\documentclass[proc]{edpsmath}
\usepackage{graphicx}
\usepackage{booktabs}

\begin{document}

\title{A 3D discrete model of the diaphragm and human trunk}

\thanks{This project was partly funded by CNRS ACINIM LePoumonVousDisJe and CNRS Inter-EPST program Bio-Informatique.}

\author{Emmanuel Promayon}
\address{TIMC-IMAG, CNRS UMR 5525, Universit\'e Joseph Fourier, Grenoble,
Institut d'Ing\'enierie de l'Information de Sant\'e, Domaine de la Merci, F-38706 La Tronche Cedex, France.
\email{Emmanuel.Promayon@imag.fr Pierre.Baconnier@imag.fr}}
\author{Pierre Baconnier}
\sameaddress{1}
\secondaddress{CHU Grenoble, Hopital Michallon, F-38700 La Tronche, France}

\begin{abstract}
In this paper, a 3D discrete model is presented to model the movements of
the trunk during breathing.
In this model, objects are represented by physical particles on their contours.
A simple notion of force generated by a linear actuator allows the model to create
forces on each particle by way of a geometrical attractor.
Tissue elasticity and contractility are modeled by local shape memory and muscular
fibers attractors.
A specific dynamic MRI study was used to build a simple trunk model comprised of by three
compartments: lungs, diaphragm and abdomen.
This model was registered on the real geometry.
Simulation results were compared qualitatively as well as quantitatively to the experimental data, in
terms of volume and geometry.
A good correlation was obtained between the model and the real data.
Thanks to this model, pathology such as hemidiaphragm paralysis can also be simulated.
\end{abstract}

\begin{resume} 
Dans cet article nous pr\'esentons un mod\`ele discret 3D permettant de mod\'eliser
les mouvements du tronc pendant la respiration.
Les objets du mod\`ele sont repr\'esent\'es par des particules physiques sur leurs contours.
Une notion simple de force induite par des actuateurs lin\'eaires permet de g\'enerer 
des forces au niveau des particules en utilisant un attracteur
g\'eom\'etrique. Les propri\'et\'es \'elastiques et contractiles d'un tissu sont ainsi mod\'elis\'ees
par des attracteurs de m\'emoire de forme locale et de fibre musculaire.
\`A partir d'une \'etude sp\'ecifique en IRM dynamique, nous avons construit un mod\`ele de tronc simplifi\'e 
comprenant trois
compartiments : les poumons, le diaphragme et l'abdomen. Ce mod\`ele est recal\'e sur la g\'eom\'etrie r\'eelle.
Nous confrontons les simulations obtenues aussi bien qualitativement que quantitativement, en terme de variation de volume et de g\'eom\'etrie. Une bonne correlation est obtenue entre le mod\`ele et les donn\'ees r\'eelles.
Gr\^ace \`a ce mod\`ele nous montrons enfin que l'on peut simuler la paralysie h\'emidiaphragmatique.
\end{resume}

\maketitle

\section*{Introduction}

\begin{figure}
\begin{center}
 \includegraphics[width=0.7\linewidth]{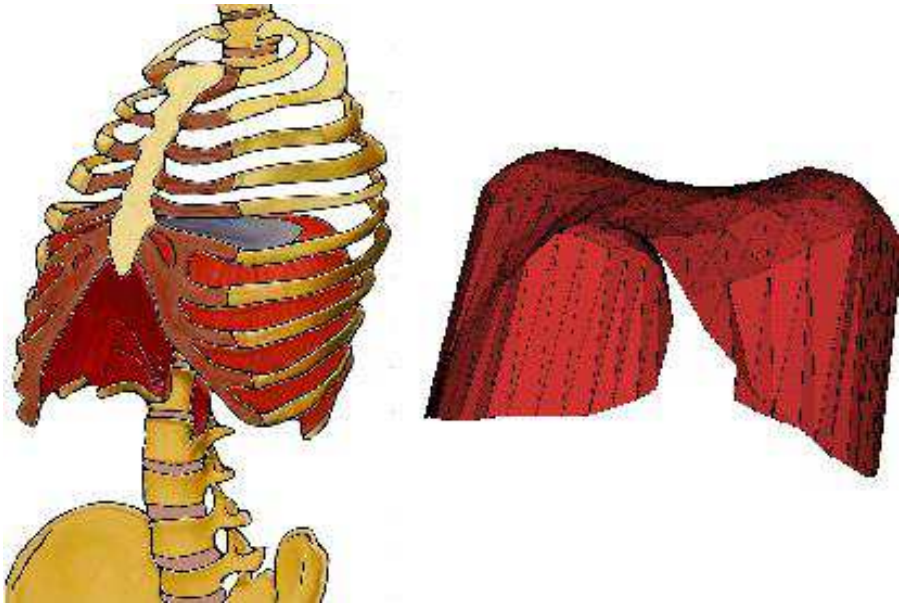}
\end{center}
\caption{\label{fig:human-trunk} Human trunk. The diaphragm and its skeleton
attachment (left). Reconstructed diaphragm surface (right).}
\end{figure}

The diaphragm has two main roles: anatomically it separates the thoracic
compartment from the abdominal compartment and physiologically it
is the main respiratory muscle. The action of this muscle is complex
and depends mainly on its size, its shape, and its attachments and
links to surrounding organs and skeleton. The human adult diaphragm
is shaped like a dome: a central tendon originates the muscular fibers.
Laterally the fibers are inserted on the 7th to the 12th ribs (see
Fig. \ref{fig:human-trunk}, left). During inspiration, the diaphragm
contracts and the abdominal content plays the role of a lever resulting
in an enlargement of the thoracic cavity. This enlargement generates
a negative pressure inside the rib cage, drawing air into the lungs.
When the diaphragm relaxes, the air is expelled, helped also by the
elasticity of the lung and the tissues lining the thoracic cavity.
The abdominal compartment can be considered as incompressible during
a given period of time (several minutes). 
Indeed, apart from a small gastric gas content, the abdominal cavity is filled with
organs of quasi constant volume (blood volume variations are
neglected) as all human tissues except lung. The stomach is commonly
isolated from the remaining digestive tract by two closed sphincters,
its gas content is then constant and considered incompressible in the
range of observed gastric pressures.

A model of the diaphragm and its surrounding structures can be used
in two simulation fields: physiology and computer assisted surgery.
It has to be geometric and kinematic as well as dynamic. If the simulated
movements are produced by the model at a sufficiently fast rate, it
can be used to predict the diaphragm and abdominal organ positions
during respiration thus being able to drive an imaging device or a conformative
radiotherapy. It is also important to be able to model specific diaphragm
pathologies, such as hemidiaphragm paralysis, as they can highly alter
the abdominal organ movements.

Physiological studies of the respiratory system classically include
volume and pressure variations. But as the diaphragm is not visible
nor easily accessible from outside the body, studying the diaphragm
deformation requires to use three dimensional medical images \cite{Whi87},
either Computerized Tomography (CT) scan or Magnetic Resonance Imaging
(MRI). Pettiaux and al \cite{PCP+97} showed that CT scan allows satisfying
3D reconstructions of the diaphragm. Cluzel and al \cite{CSC+00}
and Craighero and al \cite{CPB+05} shown similar results using MRI.
There are few works dedicated to model the diaphragm muscle. Boriek
and al. \cite{BR97} used a Finite Element Method (FEM) membrane model
to study the material behaviors but did not try to compare the model
with experimental deformations. Kinetic modeling was also proposed
in \cite{GVE+94} and \cite{PVE+92}, using geometrical change to
describe muscular actions. In respiratory physiology, the most famous
model is a compartmental model where the rib cage and abdomen form
two compartments and where an electric schema analogy is used to display
the relationships between active and passive links. However it seems
difficult to use this model to establish links between a given pathology
and some local mechanical problems and to give 3D geometric information.
Improvements of this model had recently been used to include planar
geometry information \cite{BCC+02}. Other models include computer
graphics model, such as \cite{ZCC+04} and \cite{ZCC+06}, and focus on computer graphics 3D animation
rather than physiological realism.

\section{Method}

\subsection{Available data}

\begin{figure}
\centering\includegraphics[width=0.7\linewidth]{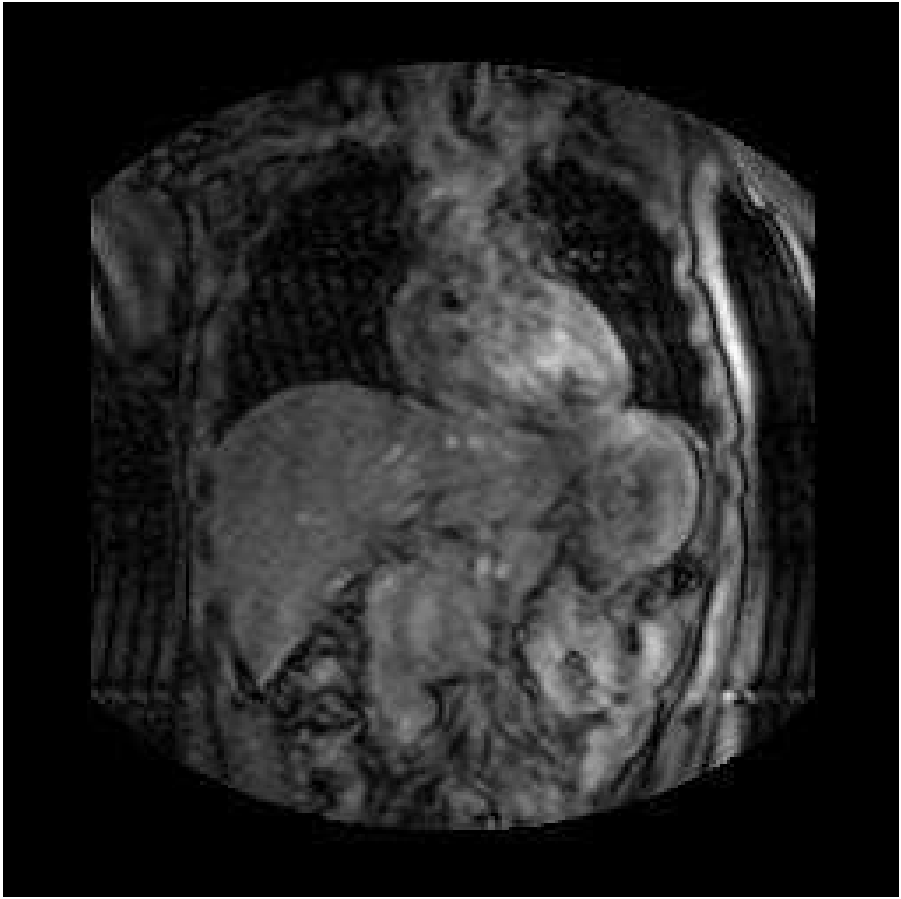}
\caption{\label{fig:An-MRI-image}An MRI image acquired using the Fast Field
Echo-Echo Planar Imaging. The resolution is 256$\times$256 pixels.
Note the variation of contrast, and the artifacts due to the acquisition
velocity.}
\end{figure}

Our current work is based on data acquired by a specific acquisition
protocol \cite{CPB+05}. A 1.5 Tesla MRI acquisition was performed
using the Fast Field Echo-Echo Planar Imaging techniques. In conventional
MRI each image is acquired in 5 seconds, has a resolution of 512$\times$512
pixels and a thickness of 1 mm. In \cite{CPB+05}, the acquisition
time was reduced to 227 ms in order to study the diaphragm deformations.
The main drawbacks of this technique are the image resolution (256$\times$256
pixels), the slice thickness (10mm) and the poor quality of the image
due to movement and reconstruction artifacts (see Fig. \ref{fig:An-MRI-image}
for an example of the acquired MRI data). Recently, using an enhanced protocol,
MR images of three ventilated subjects were acquired. The respiratory
volume and its variations were directly controlled by using artificial
ventilation. From the MRI images, a post-synchronization process made
possible the reconstruction of diaphragmatic surfaces (see Fig. \ref{fig:human-trunk},
right).

\subsection{Model}

As presented in the previous paragraph, the only available data are
poor quality MRI images (see Fig. \ref{fig:An-MRI-image}).
This yields to very strict specifications
for the geometry and paramaters registration of the model. Another
requirement for the model, as stated in the introduction, is its ability
to produce very fast and accurate simulation. Two different directions
can be taken by researchers to model human soft tissues \cite{Del98}:
the classical biomechanical approach and the computational discrete
approach. The classical biomechanical approach is mainly based on
the Continuum Mechanics  or uses its counterpart, the FEM. It offers
the advantage of being based on a strong theoretical background. There
are generally two kinds of drawbacks when one applies this method to computer
aided medical or physiological simulation: the computation time cost
and the difficulties to build complex assemblage where rigid, elastic
and active structures are interacting. Computation time can be reduced
in this context, even for material with non-linear constitutive law, by using recent
derived methods such as \cite{CDA99,DDP+01,WH04} or alternative continuous
models such as \cite{DCA99,JP03,BJ05,TBN+03}. Finite Element Analysis
is extremely interesting when one needs to understand and to know the
consequences of a local deformation on the material stress and strain. However,
in the present work, the aim is mainly to get an accurate, patient-specific
geometry and dynamics.
We need to know the consequences of the diaphragm contractions
in terms of body structure displacements and deformations, i.e. at a
higher scale than the many different materials composing the different
organs and tissues. In the FEM, the extraction of physical parameters,
such as Young modulus and Poisson ratio for linear constitutive law,
is possible by measuring isolated
tissue samples  \cite{Fun93}. \emph{In vivo} tissue characterization
is  essential because \textit{\emph{the mechanical behavior of soft
tissues can differ significantly between in-vivo and ex-vivo conditions}}.
Tissue characterization, as done in \cite{SSB+07,NMF+08}, is nevertheless
extremely difficult to perform on living tissue and/or \textit{in vivo},
notably due to the tissue accessibility, the organ movements or deformations,
and the need to sterilize the measurement devices. In this study the in-vivo
measurement of the tissues rheology is impossible as no direct access
to the organs is provided. Moreover the muscle activation function
are not anyhow available. This means that a global optimisation process
using the whole organ geometries and deformations during respiration
has to be enough to fit the geometry and the physical parameters of
the model.

Our aims are to include multiple dynamic interactions and properties,
to be able to produce real-time simulations, and to be able to fit
the model only using the available MRI data. These aims justify the
choice of a discrete approach. Previous works from the authors \cite{PBP96,PBP97},
and more recently from Zordan and al. \cite{ZCC+04,ZCC+06}, used
the same approach to build a visual simulation of the respiration.
But in both cases, the simulation were not compared to patient-specific
data nor even to experimental data. In this paper we propose to qualitatively
and quantitatively compare the results of our discrete model simulation
with the available MRI data.

\paragraph*{}
\paragraph*{}

To model living structures, we mainly need three different kinds of
components: 
\begin{itemize}
\item rigid components (to model skeleton), 
\item deformable components (to model soft tissues), 
\item and active deformable components (to model muscles). 
\end{itemize}

In our model, these components are all derived from the same principle:
a set of particles control the component surfaces, themselves organized
using triangular facets. Each particle has a position, a mass and
different properties depending on the kind of component it is part
of. Accordingly, the particles in deformable components have an elastic
property and the particles in active deformable components also have
a contractile property. In essence, this is similar to a mass-spring
network, but the elasticity is described using an original formulation,
which allow better stability and control than mass-spring network,
as shown in \cite{PBP97}.

\subsection{Dynamics}

Forces are exerted on the particles to generate displacements and deformations.
Three kind of forces are needed : 

\begin{description}
\item [{Force~field}] this kind of force is applied to all particles.
At each time the force intensity and direction is known. This kind
of force can vary depending of some mechanical or physical properties,
e.g. the mass or the velocity of the particle. The gravitional force
is an example of such a force.
\item [{Focal~force}] this is a kind of force known in intensity and direction
and applied at specific time of the simulation. For example it can
be a force applied to a particle by one of its neighbor in a particular
type of interaction. This kind of force is also used to apply boundary
conditions or to transmit user interaction in the model.
\item [{Linear~actuator~force~(LAF)}] this kind of force depends
on the internal state of the object, i.e. mainly on the geometry of
neighboring particles. The intensity and direction of this kind of
force is computed using local geometric or mechanical data and is
generally updated at each time step. A LAF is used when
a particle has to go toward an ideal position that minimizes a given
function. Therefore, LAF are used to model elasticity and contracility.
\end{description}
 We introduce a LAF in order to minimize a given energy $E$. Whenever
it is possible to define, for a given particle of 3D position $\mathbf{P}$,
a position $\mathbf{P_{minE}}$ that is known to be minimizing $E$, a LAF
can be used. Thus a LAF is simply a force that tends to minimize the
distance $|\mathbf{PP_{minE}}|$ . To express a LAF, we can use a simple expression
such as: \begin{equation}
\mathbf{F}=k_{LAF}(\mathbf{PP_{minE}})\end{equation}
 where $k_{LAF}$ is a parameter of the particle, or of a whole components. 
LAFs can thus be seen as potential forces that tends to minimize a distance. LAFs can
model any kind of forces that could be defined by a target position.
$\mathbf{P_{minE}}$ can depend on geometry or on constraints. The most important
and difficult part is to determine a correct expression for $\mathbf{P_{minE}}$,
so that it approximates a local minimum of $E$.

\paragraph*{}
\paragraph*{}

\begin{figure}
\centering\includegraphics[width=0.7\linewidth]{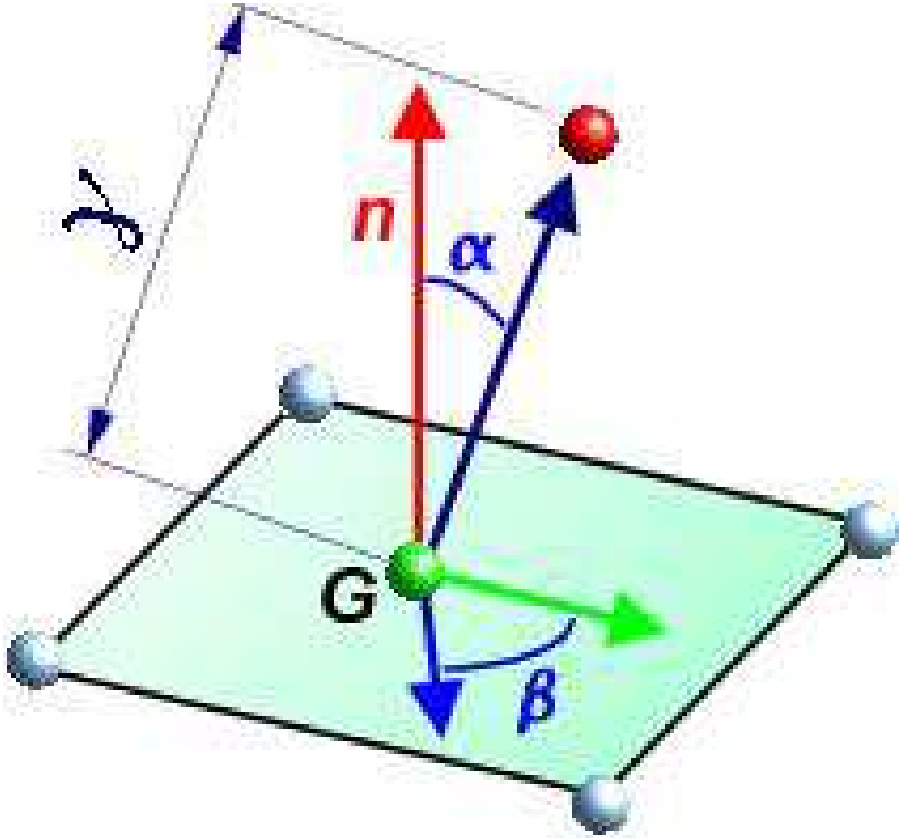}
\caption{\label{fig:Local-coordinate-system}Local coordinate system use to
define the shape function. For each particle three geometric parameters
are used: $\alpha$, $\beta$ and $\gamma$. Once determined at rest
shape, they can be use to target the position that minimize the deformation. 
The considered particle is in red, its neighbors are in blue, 
$G$ is the center of the neighbors masses, $n$ the approximated normal to the surface 
around the particle.}

\end{figure}

Spring-mass network parameters are known to be difficult to find
and adjust. Therefore, our model does not use a network of springs
to link the particles. To model the elastic property of a particle
we define a local elasticity memory \cite{PBP96}. The elastic property
of a particle is simply its ability to come back to its original geometric
configuration once deformed. To model this property each particle
has a local coordinate system defined relatively to its neighboring
particles. This local coordinate system is defined by three parameters:
two angles $\alpha$ and $\beta$ and one distance $\gamma$ \cite{PBP96},
see Fig. \ref{fig:Local-coordinate-system}. These three scalars are
initialized at the rest configuration and are called $\alpha_{0}$, $\beta_{0}$ and $\gamma_{0}$.
 At any time, if a particle position
verifies $\alpha=\alpha_{0}$ , $\beta=\beta_{0}$ , and $\gamma=\gamma_{0}$
, then the particle is at the rest configuration, thus the component
is locally undeformed. Using this local coordinate system, we can
compute at any time and for each particle a position using $\alpha_{0}$, $\beta_{0}$,
 $\gamma_{0}$ and the position of neighboring particles.
This position is ensured to locally minimize the deformation energy.
Using this position as $\mathbf{P_{minE}}$ allows us to define a LAF to minimize
this energy.

An elastic component is defined by a contour where this particular
LAF is applied to all the particles. The elasticity parameter
is the stiffness $k_{elasticity}$ used by the LAF.

\paragraph*{}
\paragraph*{}

Another LAF is used to model contractility. Once the contraction directions
(muscular fibers) on the muscular component are defined, the position $\mathbf{P_{minE}}$ 
of a particle at one extremity of a contractile components
can be simply defined as being the particle position at the other extremity
of the fiber. The LAF could then be activated by varying the $k_{contractile}=A(t).k_{muscle}$
coefficient during the simulation. $k_{muscle}$ is constant. In order
to activate the muscle contraction, $A(t)$ mimics a muscle activation
signal. $A(t)$ can take all the values between {[}0..1]. When $A(t)=1$
the activation is maximal, and when $A(t)=0$, it is null.

\paragraph*{}
\paragraph*{}

To solve the system dynamics, at each time $t$, internal and external
forces are computed, and the equation of motion are integrated, taking
into account the local and global constraints.

Note that a particle mesh can include any types of particle. For example,
it is possible to have an elastic particle with muscular neighbors.
This does not generate any computational problem. Each particle accumulates
its internal forces (elastic forces or muscular forces) and corresponding reaction forces are then distributed to its neighbors, in order to verify Newton's second law, independently
of their types.

All the geometry and physical parameter are described using the PML
language \cite{CP04}.

\subsection{Constraints and loads}

Forces are not often sufficient to model complex behaviors. Constraints
are added to maintain some conditions like non-penetrating area or
incompressibility. Our algorithm considers constraints as non-quantified
force components: they are solved using a direct projection algorithm
based on the gradient vector of the constraint function.

\paragraph*{}
\paragraph*{}

\paragraph*{\textbf{Volume preservation}}

It is possible to handle the total incompressibility of a closed contour, and
therefore to have
a tighter link with real tissues. Improvements
of the previously published method (see \cite{PBP96}) allows for
real-time computation of this particular constraint and thus for any
kind of triangulated surface.

Volume preservation is an essential property of soft tissue modeling.
The control of the volume is necessary in order to simulate both the
incompressibility of some organs and to control the volume variation
of other organs.

Consider one object described by a triangular mesh of particles at
the contour, in our model, the volume preservation constraint is applied
to all these particles. Note that the triangular mesh can also be
used for visualization. Let $n$ be the number of particles of this triangular
mesh. Let $\mathbf{P_{i}}$ denotes the positions
of the $n$ particles. Let $V_{0}$ be the initial (rest shape) volume
of the mesh and $V(\mathbf{P_{1},\cdots,P_{n}})$ a function of the particle positions that gives the
current volume value. If the volume-controled mesh is 
deformed during the simulation, our algorithm provides
a fast and efficient way to preserve the inner volume while keeping the mesh
shape similar. Let $\mathbf{\hat{P}}_{i}$, $i\in[1\cdots n]$,
be the positions of the particles before the correction due to the
volume constraint, that is to say just after the model forces
have been summed and integrated on each particles. 
Our method is able to find the displacements to apply to each particle 
in order to correct the current volume. 
In order to find these displacements, the following system has to be solved: 
\begin{equation}
\left\{ 
\begin{array}{rcl}

\mathbf{P}_{i} & = & \mathbf{\hat{P}}_{i}+\lambda\frac{\partial V}{\partial\mathbf{{\hat{P}}_{i}}},\forall i\in[1\cdots n]\\
V(\mathbf{P_{1},\cdots,P_{n}}) & = & V_{0}

\end{array}
\right.
\label{eq:vol}
\end{equation}

where $\mathbf{P}_{i}$ are the corrected positions and $\lambda$ is the unknown
scalar. $\lambda\frac{\partial V}{\partial\mathbf{{\hat{P}}_{i}}}$
is equivalent to a constrained corrective displacement that solve
the volume-preservation constraint. Solving system (\ref{eq:vol}) allows
us to directly find a solution for the volume-preservation problem.
By rearranging the equations, we can simplify system (\ref{eq:vol}) into an equation in
 $\lambda^3$ which coefficients depend only on $\mathbf{{\hat{P}}_{i}}$ and $\lambda\frac{\partial V}{\partial\mathbf{{\hat{P}}_{i}}}$.
Compared to lagrangian methods (lagrangian multiplier and minimization
algorithm), our method exactly solves the constraint and is very fast as it is mainly the 
direct solution of a third degree equation.
Note that this algorithm can also be used to control the volume variation of an object
by modifying the targeted $V_{0}$ during simulation.

\paragraph*{}
\paragraph*{}

\paragraph*{\textbf{Boundary conditions}}

On top of all forces and the volume preservation constraint, any other
boundary conditions can be applied, such as null or imposed displacement
in any direction, and imposed forces. All this kind of boundary conditions
are described using the LML language \cite{CP04}, allowing for a
dynamic change of the boundary conditions if needed during the simulation.

\subsection{Healthy and pathological diaphragm models}

\begin{figure}
\centering \includegraphics[width=0.7\linewidth]{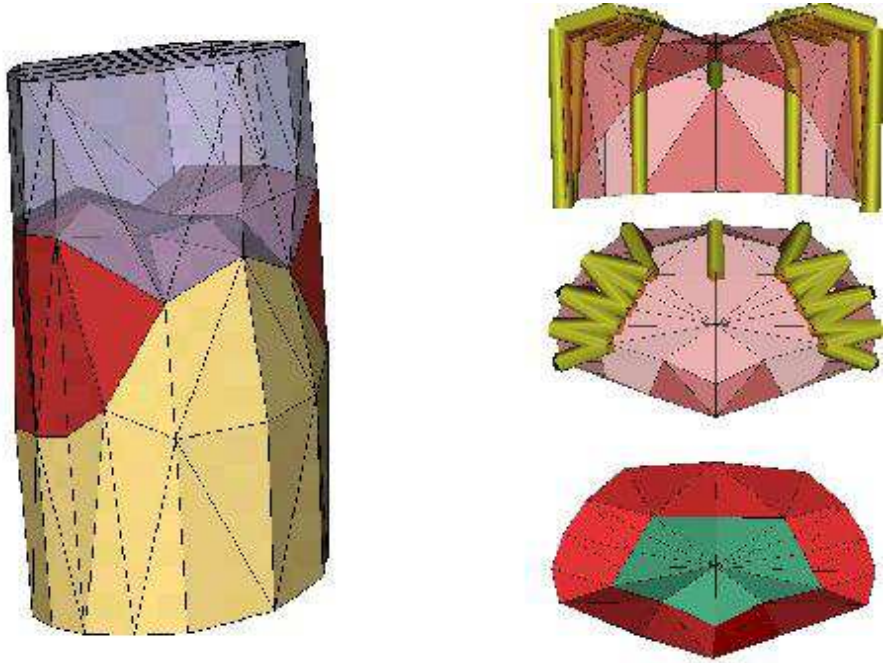} 

\caption{\label{fig:modelE} The healthy diaphragm model. The complete model
using three compartments (lungs, diaphragm and abdomen) is shown on
the left. The contraction fibers in the healthy diaphragm are represented
as cylinders (top right). The central tendon of the diaphragm is shown
on the bottom right.}

\end{figure}

Our discrete modeling framework was used to describe a simplified
human trunk (see Fig. \ref{fig:modelE}, left). It includes three
components: lungs, diaphragm and abdomen for a total of 113 particles.
The lung is a passive area, and is only modeled geometrically to monitor
the volume variation $\Delta V$ generated by the diaphragm contraction.
The diaphragm is modeled using an elastic and contractile component.
The abdomen is an elastic component. The model geometry was registered using 
an elastic matching algorithm
to the geometry segmented and reconstructed from the MRI at the beginning 
of inspiration.

The muscular fibers are defined on the model by selecting the particle
that mimic the real muscular fibers: their direction is vertical
and along the zone of apposition (see Fig. \ref{fig:modelE}, top
right). The activation function is set to mimic the physiological
signal (linear contraction for 2 seconds, important decrease for 0.5
seconds and then normal decrease for 2 seconds). To model the central
tendon, we set $k_{elasticity}$ of the top central part of the diaphragm
as being twice as rigid as the other diaphragm areas (see Fig. \ref{fig:modelE},
bottom right).

As the abdominal compartment can be considered as incompressible,
the boundary conditions essentially consist in maintaining the incompressible
constraint on the mesh defined by the diaphragm and the abdomen walls.
A null displacement boundary condition is imposed to some particles
to model the spine, the pelvis, and the top of the lung.

\begin{figure}
\centering \includegraphics[width=0.7\linewidth]{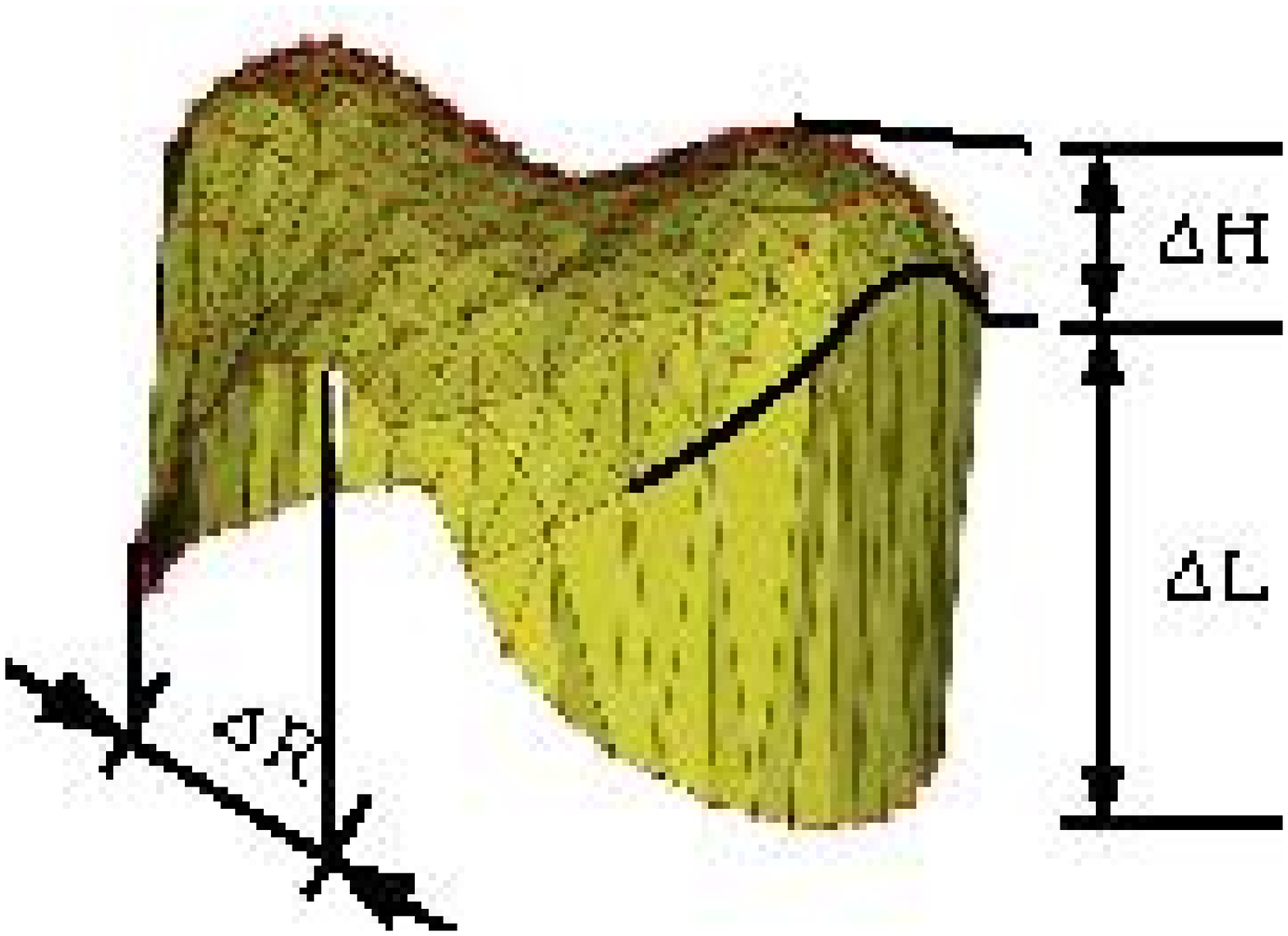} 

\caption{\label{fig:measurements} Comparison measurements. Deformations are
measured in the zone of apposition ($\Delta L$), the height of the
diaphragmatic domes ($\Delta H$) and the transverse direction ($\Delta R$).}

\end{figure}

The model was compared with the experimental data by studying the
deformation during a respiratory cycle in different directions (Fig.
\ref{fig:measurements}). We compared deformation in terms of the
variations of, from the most significant to the least significant:
pulmonary volume ($\Delta V$), apposition zone length ($\Delta L$),
height of the diaphragmatic domes ($\Delta H$), and transverse length
($\Delta R$).

We also simulated a pathological condition: an hemidiaphragm paralysis.
This was obtained by inactivating all fibers of the same side of the
diaphragm. An additional modification was needed: the pathological
model has to include the long term effect of the paralysis, namely
the elasticity loss of the paralysed hemidiaphragm. We thus set a
different value of the elasticity parameter $k_{elasticity}$ for
half of the diaphragm tissue. This pathological situation is known
to lead to many ventilatory impairments among which a drastic decrease
of inspiratory muscle efficiency, inducing a decrease of tidal volume
(total volume displacement of each breath) and a paradoxical upward
(\char`\"{}expiratory\char`\"{}) movement of the paralysed hemidiaphragm
during inspiration.

\subsection{Estimation of the model parameters}

In order to estimate the model parameters, the only available data were the
pulmonary volume variation $\Delta V$. The main advantage of our
model is its very fast computation time and its reduced number of
parameters. In the trunk model, only two parameters are to be estimated:
$k_{elasticity}$ and $k_{muscle}$. An optimization algorithm based
on an \textquotedblleft{}analysis by synthesis\textquotedblright{}
strategy was elaborated. It consisted in a four step loop: (1) assume
a given set of parameters, (2) build and simulate a respiratory cycle
using the model, (3) compare the provided simulations with the respiratory volume
measurements in the least square sense, (4) from this comparison deduce
better values of parameters in order to improve the simulation/measurement
fit. This loop was continued until the comparison carried out in (3)
gives satisfactory results.

\begin{center}
\begin{table}
\begin{tabular}{@{}lcccccccc@{}} \toprule
 \multicolumn{6}{r}{Volume and measurements} \\
\cmidrule(r){2-9}
& End expiratory & End inspiratory &
 $\Delta V$ & $\Delta V$ & $\Delta L$ & $\Delta L$ &
 $\Delta H$  & $\Delta R$  \\
& vol. (ml) &  vol. (ml) & (ml) & (\%) & (mm) & (\%) & (mm) & (mm) \\ \midrule
Experimental data &  4760 &  5226 & 466 & 9.78 & 9.13 & 6.13 &  2.00 & 0.00 \\
Model 		  &  4759 &  5197 & 438 & 9.20 & 9.73 & 5.98 & 13.30 & 8.80 \\
\% error	  & -0.02 & -0.55 &  -6 &      & 6.57 &      &       &      \\ \bottomrule
\end{tabular}
\caption{\label{table:comparison} Qualitative comparisons between real data and
model.}
\end{table}
\end{center}

\section{Results}

\begin{figure}
\centering \includegraphics[width=0.7\linewidth]{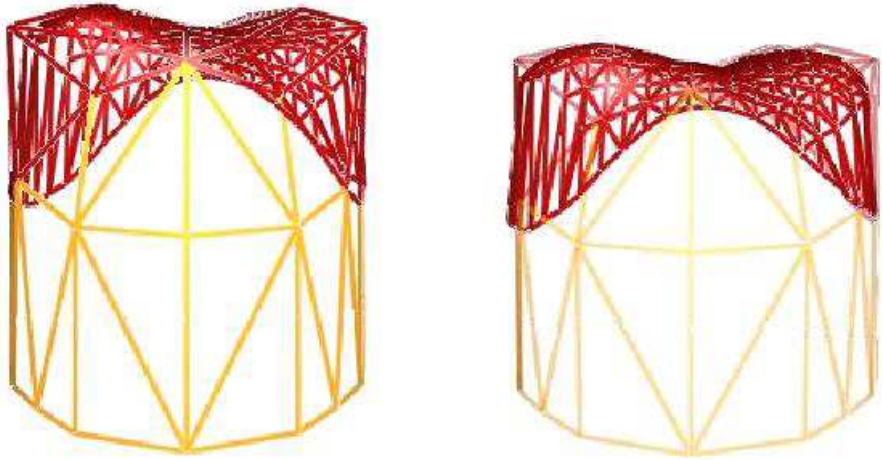} 

\caption{\label{fig:shape-matching} Comparisons of the model diaphragm geometry
(coarse mesh) and reconstructed diaphragmatic surface. The model was
initially deformed to match the reconstructed shape at beginning of
the respiratory cycle (end expiratory position) (left). After the
simulation of a respiratory cycle, the simulated deformation are superimposed
with reconstructed diaphragmatic surface at the end of inspiration
(right).}

\end{figure}

The simulation of a complete respiratory cycle only takes 1.50 seconds
on a Pentium Xeon 5140 at 2.33 Ghz, i.e. a frame rate of approximately
3000fps. Comparisons between the model and the real data are presented
in Table 1. Qualitative geometry comparisons were also made between
the surface of the diaphragm in the model and the reconstructed diaphragmatic
surface at the end of inspiration (see Fig. \ref{fig:shape-matching}).

\begin{figure}
\centering \includegraphics[width=0.7\linewidth]{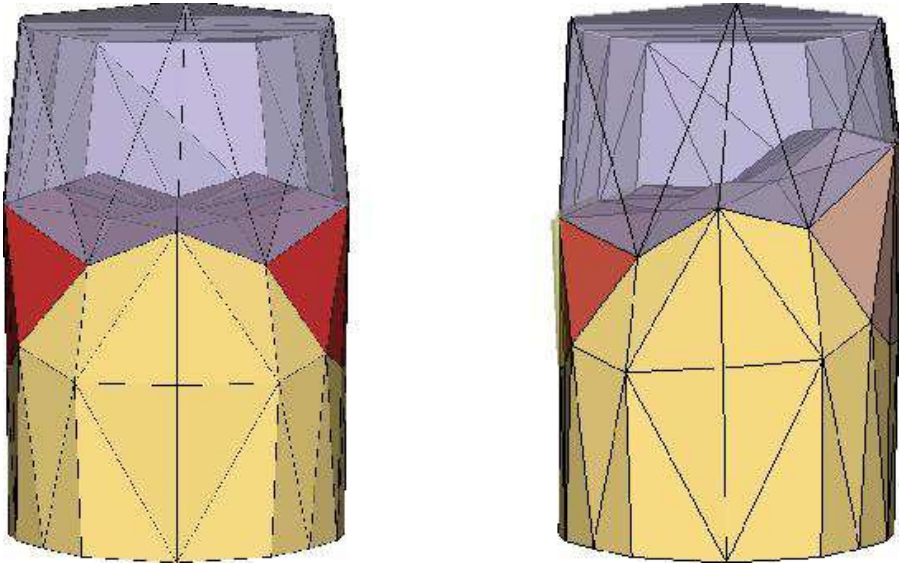} 

\caption{\label{fig:hemidiaphragm} Comparison between a healthy diaphragm
model and hemidiaphragm paralysis model using the same activation
function at end inspiration
(during exercise breathing).}

\end{figure}

The model is able to simulate the hemidiaphragm paralysis pathology.
Comparisons between healthy and pathological diaphragm can easily
be observed in 3D (see Fig. \ref{fig:hemidiaphragm}).

\begin{figure}
\centering \includegraphics[width=0.7\linewidth]{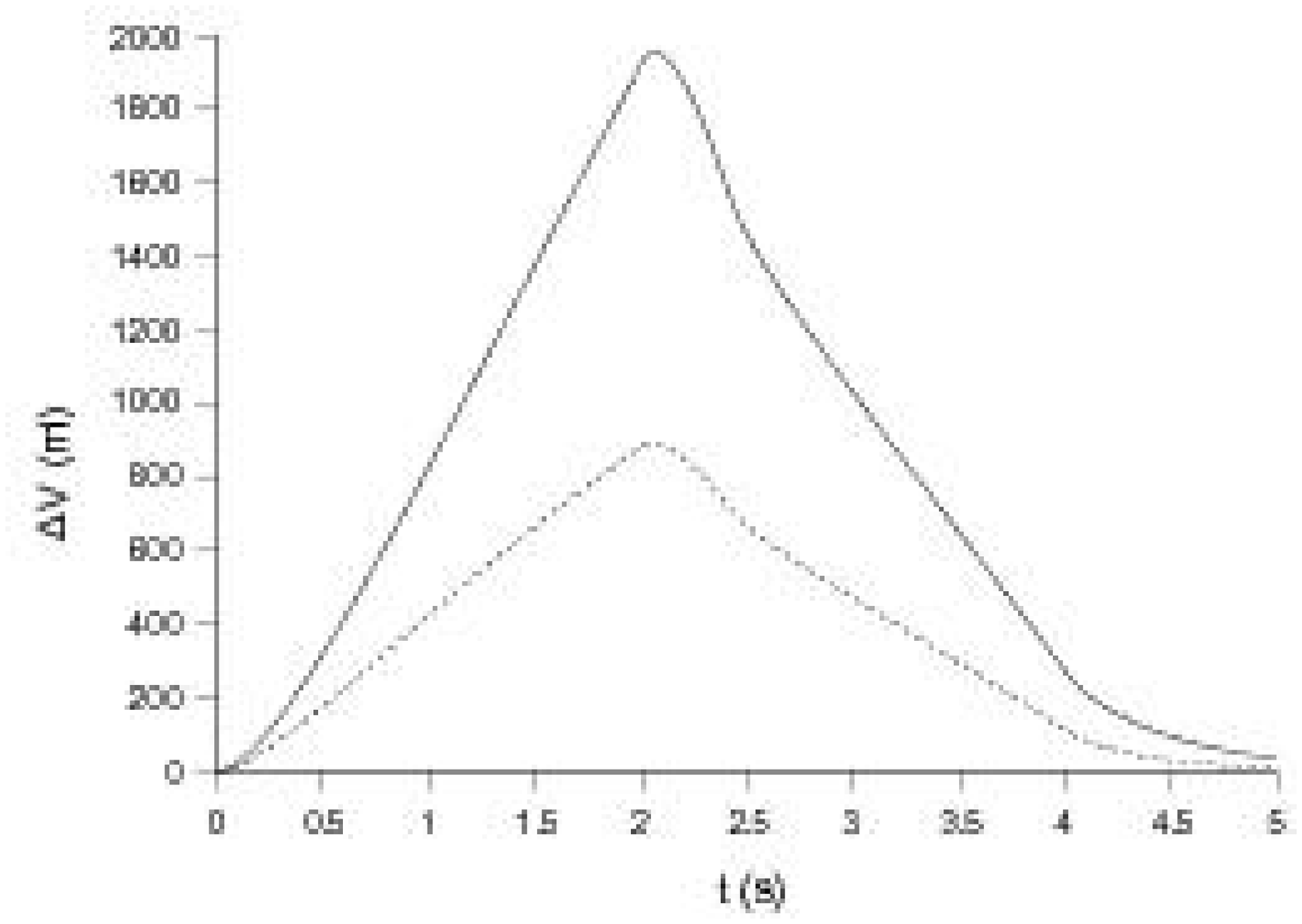} 

\caption{\label{fig:hemidiaphragm-volume} Comparison of the volume variation
during exercise breathing generated by the healthy diaphragm model
(continuous line) and the hemidiaphragm paralysis model (dashed line).}

\end{figure}

We also can see significant differences between lung volume displacements
(see Fig. \ref{fig:hemidiaphragm-volume}).

\section{Discussion}

\subsection{Healthy diaphragm}

The model was able to reproduce an accurate volume variation $\Delta V$
and piston-like deformation $\Delta L$. The model deformation compared
to real deformation measured on a subject was as well qualitatively
satisfying. Another very important point, especially when considering the application
of the method in computer assisted medical intervention and physiological
studies, is the fast computation time: the simulation is about four times faster
than the respiratory cycle it is simulating. This result
leaves some space for the improvement and enhancement of our model.

On the other side the simulation are far from correct when $\Delta H$
and $\Delta R$ are compared. These differences are probably due to
an over simplification in the discretization of the diaphragmatic
zone and to the model itself, which does not take contact and friction
into account.

Although discrete and very simple, this model efficiently reproduced
the complex movements of breathing. The major drawback of this model
is that being discrete, it is not possible to compute the extract
strain and stress on the different components. As these values on
the \textit{in vivo} diaphragm are not obtainable by any technique,
the choice of a continuous model does not seem to be crucial.

\subsection{Hemidiaphragm}

We observed all the clinical consequences of the simulated pathology.
The paradoxical behaviour of the paralysed hemidiaphragm (upward displacement
during inspiration) is evidenced on the 3D simulation (Fig. \ref{fig:hemidiaphragm},
right). The difference between volume displacements in the healthy
and pathological diaphragm amounts to what is typically measured (a
50\% shortening, T. Similowski, personal communication) in clinical
results (Fig. \ref{fig:hemidiaphragm-volume}).

\subsection{Future works}

\begin{figure}
\centering\includegraphics[width=0.7\linewidth]{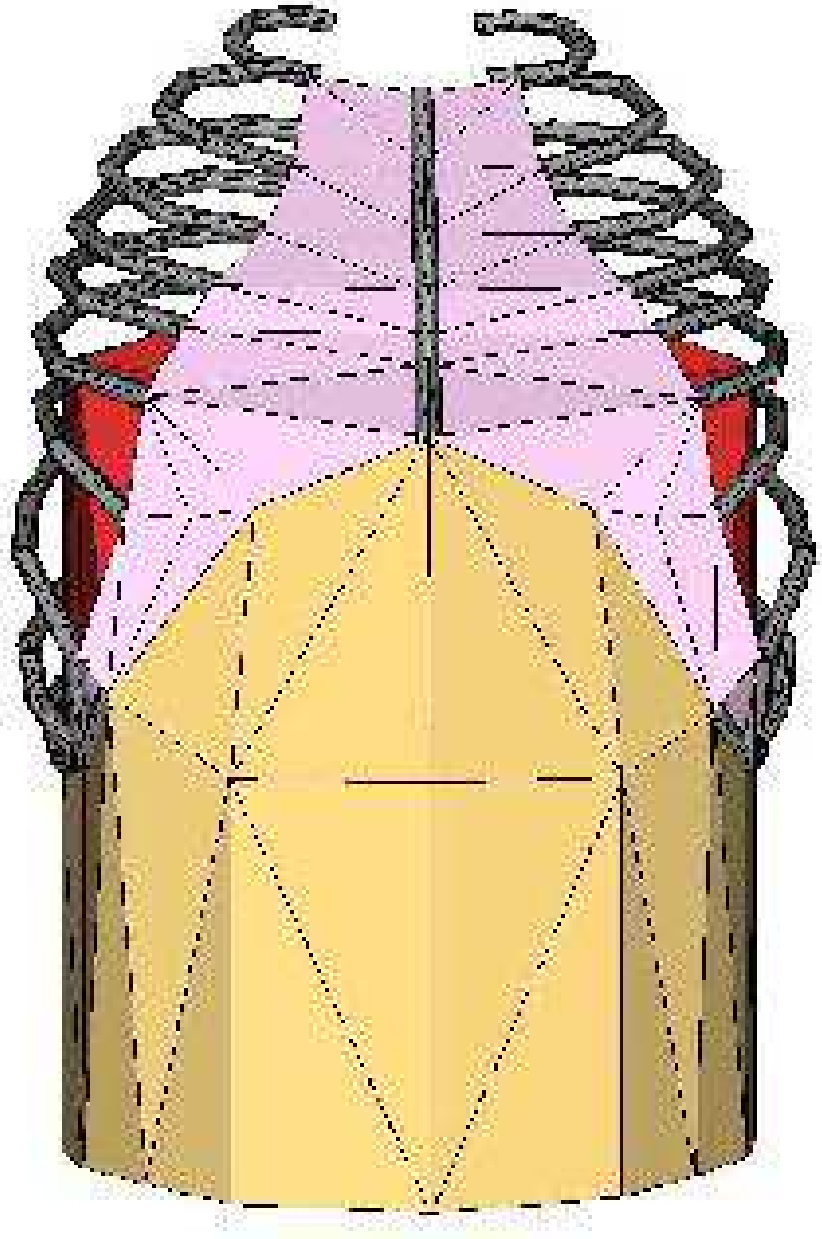}

\caption{\label{fig:Advanced-model-of}Advanced model of the human trunk including
one solid body per rib with elastic links to model cartilaginous tissues.}

\end{figure}

Once the $k_{elasticity}$ parameter is set for a given subject, the
main advantage of this optimization technique is that we can directly
and quickly adjust $k_{muscle}$ depending on $\Delta V$. This can
lead to a real-time prediction of the diaphragm position during breathing,
considering only one medical image at the beginning from which the
diaphragm geometry can be registred. This could be used for example
during conformative radiotherapy. Future works on the diaphragm model
will include testing and validating other breathing situations and
comparisons with other subject data. To overcome the differences noted
for $\Delta H$ and $\Delta R$ we also plan to add the rib cage and
its cartilage components (this work has just started, see Fig. \ref{fig:Advanced-model-of}).

\section*{Acknowledgement}

\begin{acknowledgement} The authors wish to thank L. Gaillard for
her contribution to this work, and T. Similowski for suggesting the
simulation of the hemidiaphragm paralysis. \end{acknowledgement}

\bibliography{ESAIM2008}
 \bibliographystyle{plain}
\end{document}